\documentclass[11pt,twoside]{article}
\usepackage{amsmath,amsthm, amsfonts,multirow}
\usepackage{graphicx,epsfig,natbib,epstopdf}
\usepackage{CS18}
\usepackage{hyperref}
\hypersetup{colorlinks=true,citecolor=cyan,urlcolor=cyan,linkcolor=blue}

\newcommand{\planss}{Planet. Space Sci.}

\begin{document}
%
%
%
\title{Numerical aspects of 3D stellar winds}
%
%
\author{A. Strugarek$^{1,2}$, A. S. Brun$^{2}$, S. P. Matt$^{3}$,
  V. Reville$^{2}$}
\affil{$^1$D\'epartement de physique, Universit\'e de Montr\'eal, C.P. 6128 Succ. Centre-Ville, Montr\'eal, QC H3C-3J7, Canada}
\affil{$^2$Laboratoire AIM Paris-Saclay, CEA/Irfu Universit\'e Paris-Diderot CNRS/INSU, F-91191 Gif-sur-Yvette.}
\affil{$^3$Astrophysics group, School of Physics, University of Exeter, Stocker Road, Exeter EX4 4QL, UK}
\begin{abstract}
%
%
This paper explores and compares the pitfalls of modelling the
three-dimensional wind of a spherical star with a cartesian
grid. Several numerical methods are compared, using either uniform and
stretched grid or adaptative mesh refinement (AMR). An additional
numerical complication is added, when an orbiting planet is considered.
In this case a rotating frame is added to
the model such that the orbiting planet is at rest in the frame of
work. The three-dimensional
simulations are systematically compared to an equivalent
two-dimensional, axisymmetric simulation. The comparative study
presented here suggests to limit the rotation rate of the rotating
frame below the rotating frame of the star and provides guidelines for
further three-dimensional
modelling of stellar winds in the context of close-in star-planet
interactions.
\end{abstract}
%
%
%
%
%

\section{Introduction}
\label{sec:introduction}

Magnetized stellar winds have long been recognized as the major source of
angular momentum extraction in main sequence stars
\citep{Parker:1958dn,Weber:1967kx,Mestel:1968aa}. In order to reliably
assess the stellar wind torque, the acceleration
profile and the magnetic field geometry of the wind are needed. It was recently
demonstrated that, in particular, complex magnetic topologies
of cool stars could
significantly alter the torque \citep[see,
\textit{e.g.}][]{Cohen:2014et,Reville:2014ud} compared to more
simple topologies. Three dimensional
numerical simulations provide today a reliable way to compute, in a
dynamically self-consistent way, the torque arising from stellar wind
with complex magnetic fields for a large variety of stars.
However, no consensual parametrization of fully three-dimensional,
non-axisymmetric stellar wind torques has yet been proposed in the
literature.

Furthermore, the growing number of know exoplanets triggered renewed
interest in the
recent years in the interactions existing between star and
close-in planets \citep[for a recent review, see][in this
volume]{Lanza:2014aa}. In particular, close-in planets can
magnetically
interact with their host, which leads 
to a direct transfer of angular momentum due to a
magnetic link between the two objects \citep[among numerous other
effects as well, see, \textit{e.g.},][and references there
in]{Cuntz:2000ef,Zarka:2007fo,Scharf:2010ab,Vidotto:2014kk}. Several
analytical 
studies \citep[\textit{e.g.},][and references
therein]{Lanza:2009fp,Laine:2011jt}  have been pursued in the past
years to better constrain our
understanding of this angular momentum transfer. In a
recent work, \citet{Strugarek:2014uh}
explored the efficiency of the angular momentum transfer as a function of the relative
position of the orbiting planet in the stellar wind and of the topology
of the planetary field with 2.5D simulations. In order to
validate the trends they found, 3D numerical simulations taking into
account the adequate geometry of the problem are needed
\citep[see][for an example of such global modelling]{Cohen:2009dw}.

We report here an ongoing effort in developing
magnetohydrodynamics (MHD) simulations of the stellar winds of cool
stars in three dimensions. We consider one-fluid and
ideal models of stellar winds which are very simple compared to
more recent solar wind models \citep[see,
\textit{e.g.},][]{Oran:2013aa,Sokolov:2013ki}. However, they inherit
important conservation properties from their 2.5D counterparts
\citep[see][and section \ref{sec:rotating-frame-1}]{Strugarek:2012th}.
We show in this work that ensuring such conservation
properties is mandatory to derive physically meaningful global trends
from grids of numerical simulations. By such, they are thus
of particular interest for our understanding of the gyro-chronology of
cool stars. In addition, we focus here on the numerical difficulties
associated with a rotating frame, anticipating
eventual star-planet interactions studies with such stellar wind models. 

\section{Modelling stellar winds}
\label{sec:modell-stell-winds}

Following the preliminary work in 2.5D axisymmetric geometry described in
\citep{Strugarek:2014uh}, we adapted our stellar wind model to a 3D cartesian
geometry. We implemented the same ``3-layer'' boundary conditions to
improve the conservation properties of our numerical
solution. We use the PLUTO code \citep{Mignone:2007iw} which solves the following set of ideal MHD equations:
\begin{eqnarray}
  \label{eq:mass_consrv_pluto}
  \partial_t \rho + \boldsymbol{\nabla}\cdot(\rho \mathbf{v}) &=& 0 \, \\
  \label{eq:mom_consrv_pluto}
  \partial_t\mathbf{v} +
  \mathbf{v}\cdot\boldsymbol{\nabla}\mathbf{v}+\frac{1}{\rho}\boldsymbol{\nabla} P
  +\frac{1}{\rho}\mathbf{B}\times\boldsymbol{\nabla}\times\mathbf{B}
  &=& \mathbf{a} \, ,
  \\
  \label{eq:ener_consrv_pluto}
  \partial_t P +\mathbf{v}\cdot\boldsymbol{\nabla} P + \rho
  c_s^2\boldsymbol{\nabla}\cdot\mathbf{v} &=& 0 \, ,\\
  \label{eq:induction_pluto}
  \partial_t \mathbf{B} - \boldsymbol{\nabla}\times\left(\mathbf{v}\times\mathbf{B}\right)
  &=& 0 \, ,
\end{eqnarray}
where $\rho$ is the plasma density, $\mathbf{v}$ its velocity, $P$ the gas
pressure, $\mathbf{B}$ the magnetic field, and $\mathbf{a}$ 
is composed of gravitational acceleration (which is time-independent)
and the Coriolis and centrifugal forces of a rotating frame
$\Omega_{0}$. The sound speed is given by $c_s=\sqrt{\gamma\,P/\rho}$, with
$\gamma$ the adiabatic exponent. We use an
ideal gas equation of state
\begin{equation}
  \label{eq:EOS}
  \rho\varepsilon = P/\left(\gamma-1\right)\, ,
\end{equation}
where $\varepsilon$ is the specific internal energy. We use an \textit{hll}
solver combined with a \textit{minmod} limiter. A
second-order Runge-Kutta is used for the time evolution, resulting in
an overall second-order accurate numerical method. The solenoidality
of the magnetic field is ensured with a constrained transport
method in the static grid version of the model, and with Powell's
\textit{eight waves} method in the AMR version
\citep[see][]{Mignone:2012aa}. We refer the interested reader to
\citep{Mignone:2007iw} for
an extensive description of the various methods that PLUTO offers.

The structure of the wind directly depends on three velocity ratios defined at the surface
of the star \citep[see, \textit{e.g.},][]{Matt:2012ib}, and on the
ratio of specific heats $\gamma$. The three 
characteristic velocities are the sound speed $c_{s}$,
the Alfv\'en speed $v_A=B_{\star}/\sqrt{4\pi\rho_{\star}}$ (where
$B_{\star}$ is the magnetic field strength at the stellar
equator) and the rotation speed 
$v_{\rm rot}$ (in this work, the
star is considered to rotate as a solid body). Their ratios to the escape velocity $v_{\rm
  esc}=\sqrt{2GM_\star/R_\star}$ (with $M_{\star}$ the stellar mass
and $R_{\star}$ the stellar radius) at the stellar surface then define a
unique stellar wind solution. We choose for this study the same
parameters as in \citet{Strugarek:2014uh}, which we report in table
\ref{tab:tab_params}. We also compute the rotation rates associated
with these velocities at the surface of the star and deduce the
equivalent orbital radius of a virtual planet (for a characteristic velocity
$V$, the equivalent normalized orbital radius $r_{orb}/R_{\star}$ can be approximated by
$(GM_{\star}/R_{\star}V^{2})^{1/3}$, see Equation \ref{eq:omega_rot}).

\begin{table}[htbp]
  \caption{Fiducial stellar wind parameters}
  \label{tab:tab_params}
  \centering
  \begin{tabular}[htbp]{ccc}
    \hline 
    Parameter & Value & Equivalent
    $r_{orb}/R_{\star}$ \\
    \hline
    $\gamma$ & 1.05 & ... \\
    $c_{s}/v_{\rm esc}$ & 0.2599 & 1.95 \\
    $v_{A}/v_{\rm esc}$ & 0.3183 & 1.7 \\
    $v_{\rm rot}/v_{\rm esc}$ & 0.00303  & 38 \\
    \hline
  \end{tabular}
\end{table}

We intend to ultimately use our stellar wind model to study global close-in star-planet
interactions in 3D. We choose a cartesian
grid to avoid any future numerical issues that would be associated
with very small grid cells at the stellar surface when using a
curvilinear coordinate system with structured grids. 
The star is located at the
center of our three-dimensional grid. In this work we consider two
different static grid sizes to model stellar
winds, the higher resolution being hereafter denoted HR. We also show
one preliminary simulation using the AMR version of the pluto code. In
the static version, the cube
$[-1.5\,R_{\star}, 1.5\,R_{\star}]^{3}$ centered on
the star is always uniformly discretized, and stretched grids are use in the
three directions to grid the remaining of the domain up to $20\, R_{\star}$ from
the star. The discretization is identical in the three dimensions.

In order to include a planet in such a stellar wind simulation, one
can solve the MHD equation in a rotating frame rotating at the orbital
rotation rate of the planet. The planet is then nicely at rest in the frame
of the grid, and the stellar rotation rate needs just to be modified
accordingly. Considering circular Keplerian orbits and neglecting the
orbital motion of the star, the orbital
rotation rate --that we use as the rotation rate of the
rotating frame-- of a planet located at $r_{orb}$ is given by
\begin{equation}
  \label{eq:omega_rot}
  \Omega_{0} = \Omega_{P} = \sqrt{\frac{GM_{\star}}{r_{orb}^{3}}}\, .
\end{equation}
In the following, even though we do not include any planet in the
simulations yet, we label the various rotating frames we considered
(listed in Table \ref{tab:tab_cases}) by
their equivalent orbital radius $r_{orb}$ of the virtual planet. We
express it in terms of breakup rotation rate
$\Omega_{b}=\left(GM_{\star}/R_{\star}^{3}\right)^{1/2}$.

\begin{table}[htbp]
  \caption{Parameters of the stellar wind cases}
  \label{tab:tab_cases}
  \centering
  \begin{tabular}{lccc}
    \hline
    Case & Resolution & $r_{orb}/R_{\star}$ & $\Omega_{0}/\Omega_{b}$ \\
    \hline
    1 & 225$^{3}$  & $\infty$ & 0 \\
    2 (HR) & 449$^{3}$  & $\infty$ & 0 \\
    3 (AMR) & 1920$^{3}$ & $\infty$ & 0 \\
    4 & 225$^{3}$  & 50 & 0.002 \\
    5 & 225$^{3}$  & 10 & 0.022 \\
    6 & 225$^{3}$  & 3  & 0.136 \\
    7 (HR) & 449$^{3}$ & 3  & 0.136 \\
    \hline
  \end{tabular}
\end{table}

\section{Global properties of the modelled winds}
\label{sec:comp-betw-diff}

\begin{figure}[htbp]
  \centering
  \includegraphics[width=0.49\linewidth]{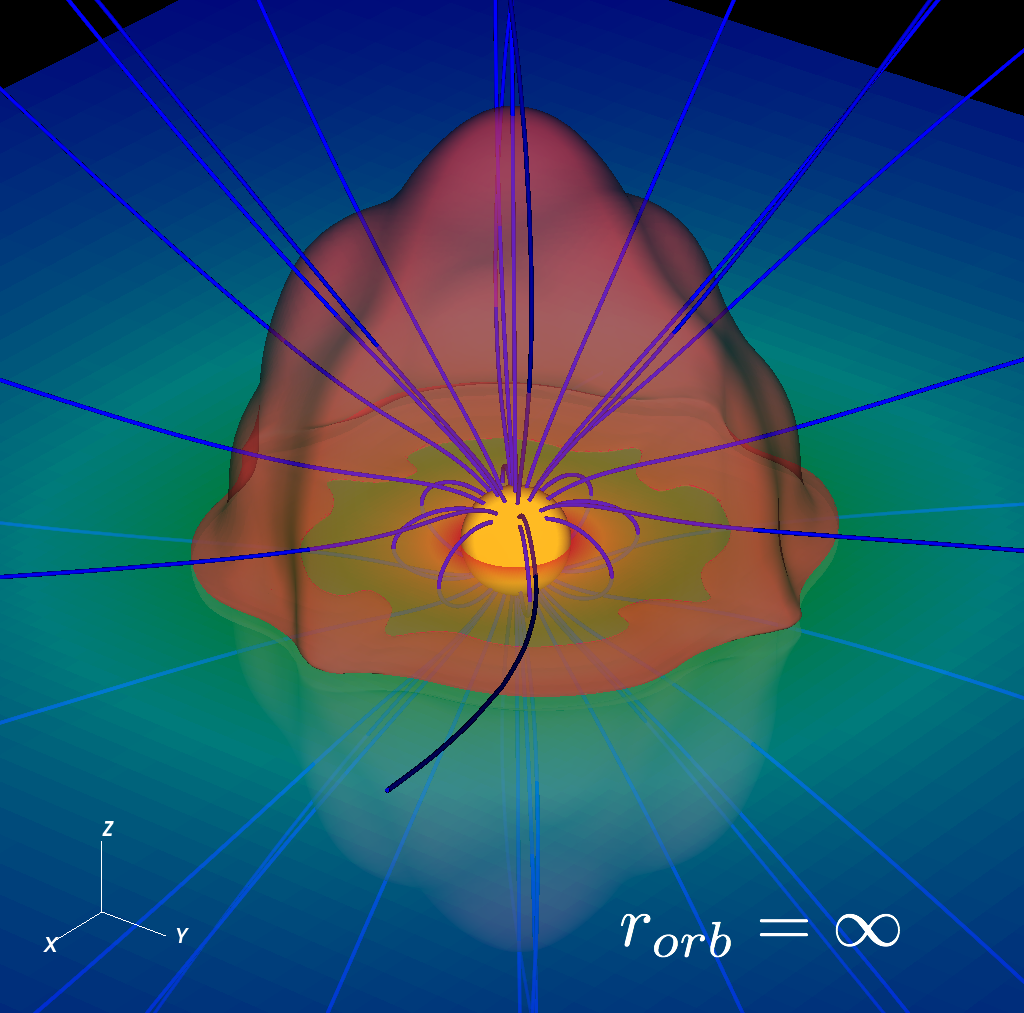}
  \includegraphics[width=0.49\linewidth]{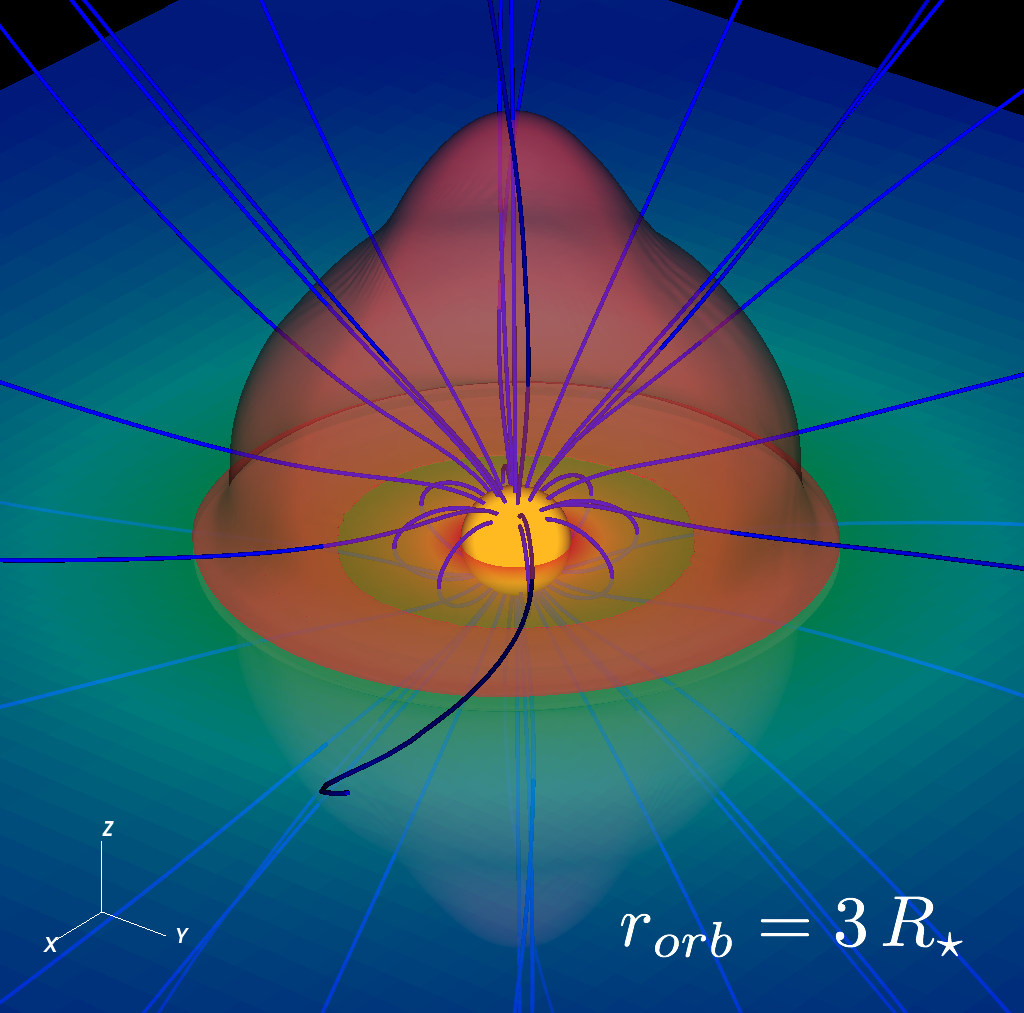}
  \includegraphics[width=0.49\linewidth]{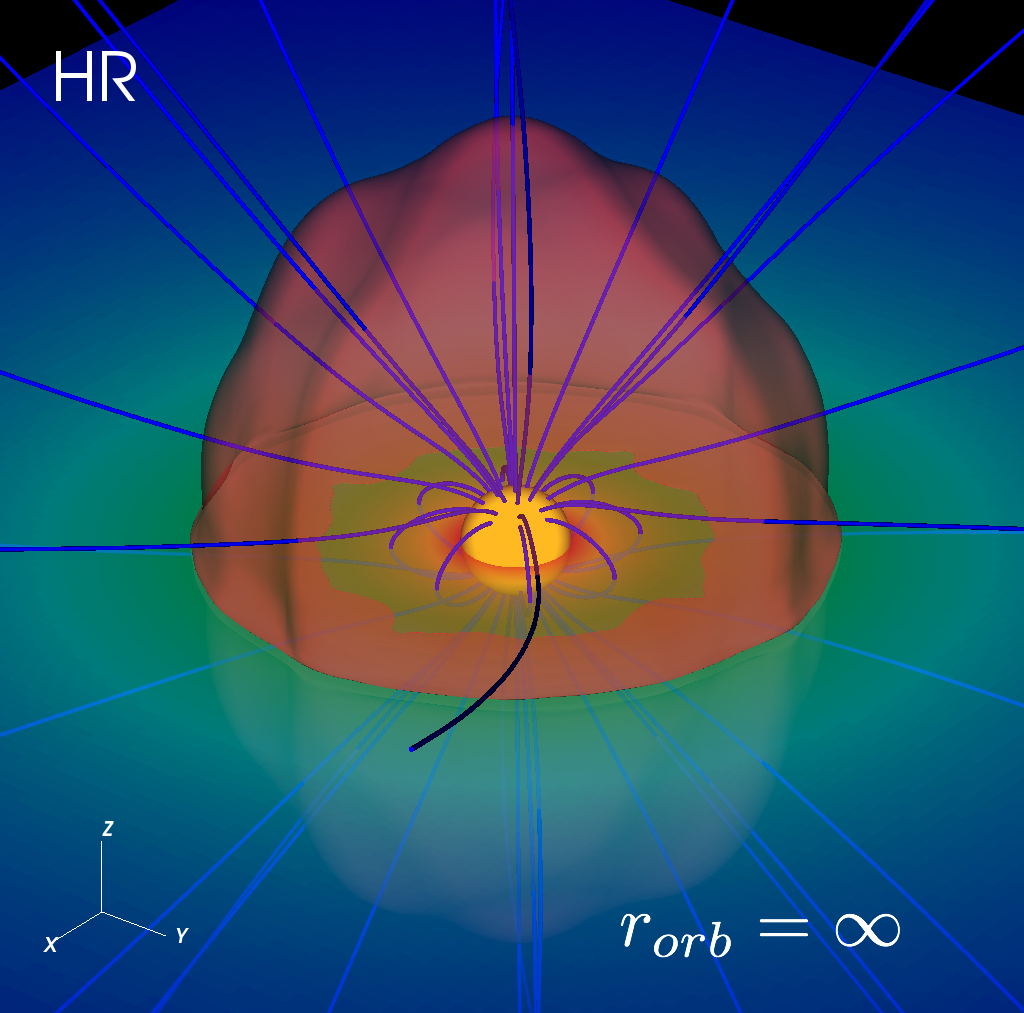}
  \includegraphics[width=0.49\linewidth]{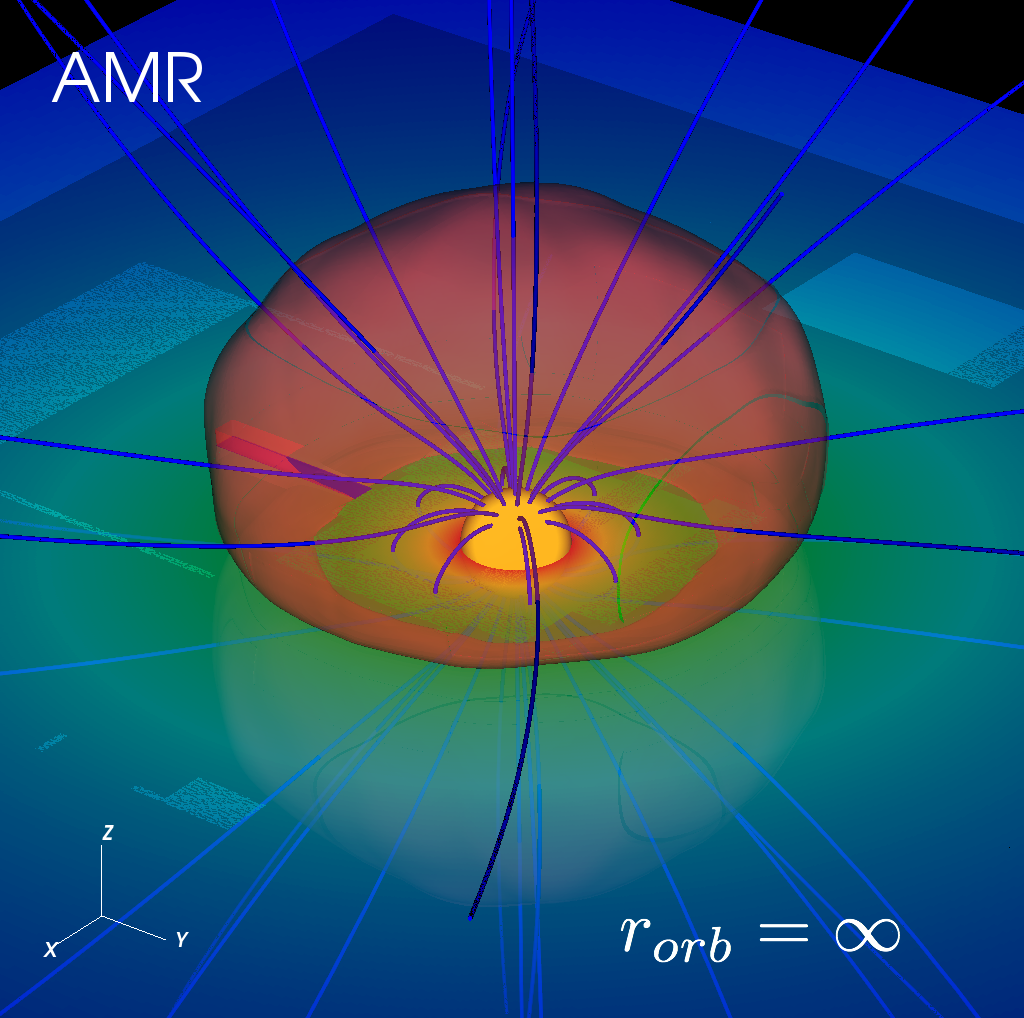}
  \caption{3D renderings of the modelled stellar winds. The upper
    panels show cases 1 and 6, and the bottom panels cases 2 and
    3 (see table \ref{tab:tab_cases}). The stellar
    boundary is labelled by the orange sphere. The magnetic field
    lines are shown in blue and the Aflv\'en surface in transparent
    red. The density on the equatorial plane in shown in logarithmic
    scale, with the same color map on all the panels.}
  \label{fig:3D}
\end{figure}

We first illustrate our modelled stellar wind with three-dimensional
visualizations of the cases 1, 2, 3, and 6 (see table \ref{tab:tab_cases})
in Figure 
\ref{fig:3D}. In all the figures presented throughout this paper, the
results have been transformed to the inertial frame
to adequately compare the various cases.

Interestingly, the addition of a rotating frame
seems at first glance
to regularize the solution: the shape of the Alfv\'en surface (where
the wind speed equals the local Alfv\'en speed) in case
1 (upper left panel) shows some non-axisymmetric features due to our cartesian grid
whereas in case 6 (upper right panel, $r_{orb}=3\, R_{\star}$) its looks perfectly
axisymmetric. Nevertheless, despite
this apparent regularization, the rotating frame induces significant
(and non-axisymmetric) deviations in the stellar wind solution that could be
problematic in the context of star-planet interactions models. We
detail and quantify these deviations in the following sections. Higher
resolution in the case with no rotating frame (lower panels, HR and
AMR cases) clearly tend to reduce the non-axisymmetric aspect of the
Alfv\'en surface.

\subsection{Mass and angular momentum loss rates}
\label{sec:resolution-effect}

\begin{figure}[htbp]
  \centering
  \includegraphics[width=0.49\linewidth]{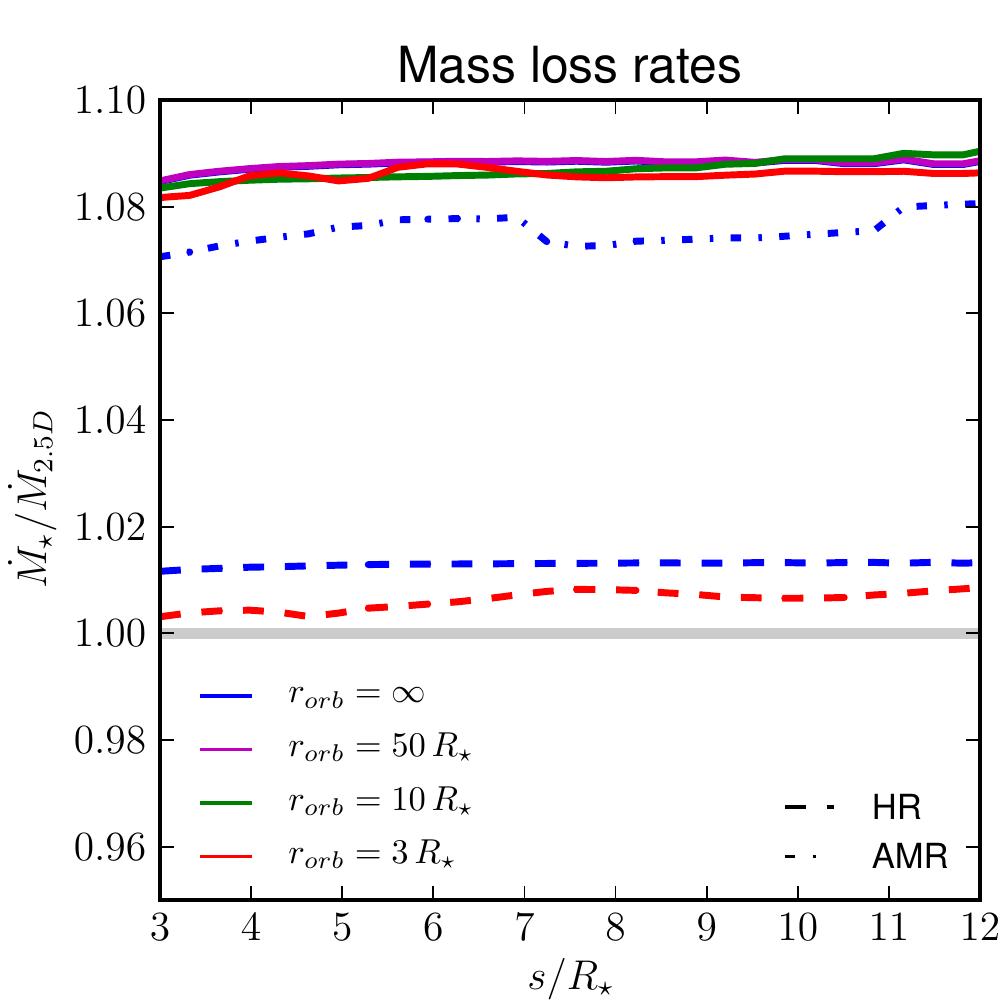}
  \includegraphics[width=0.49\linewidth]{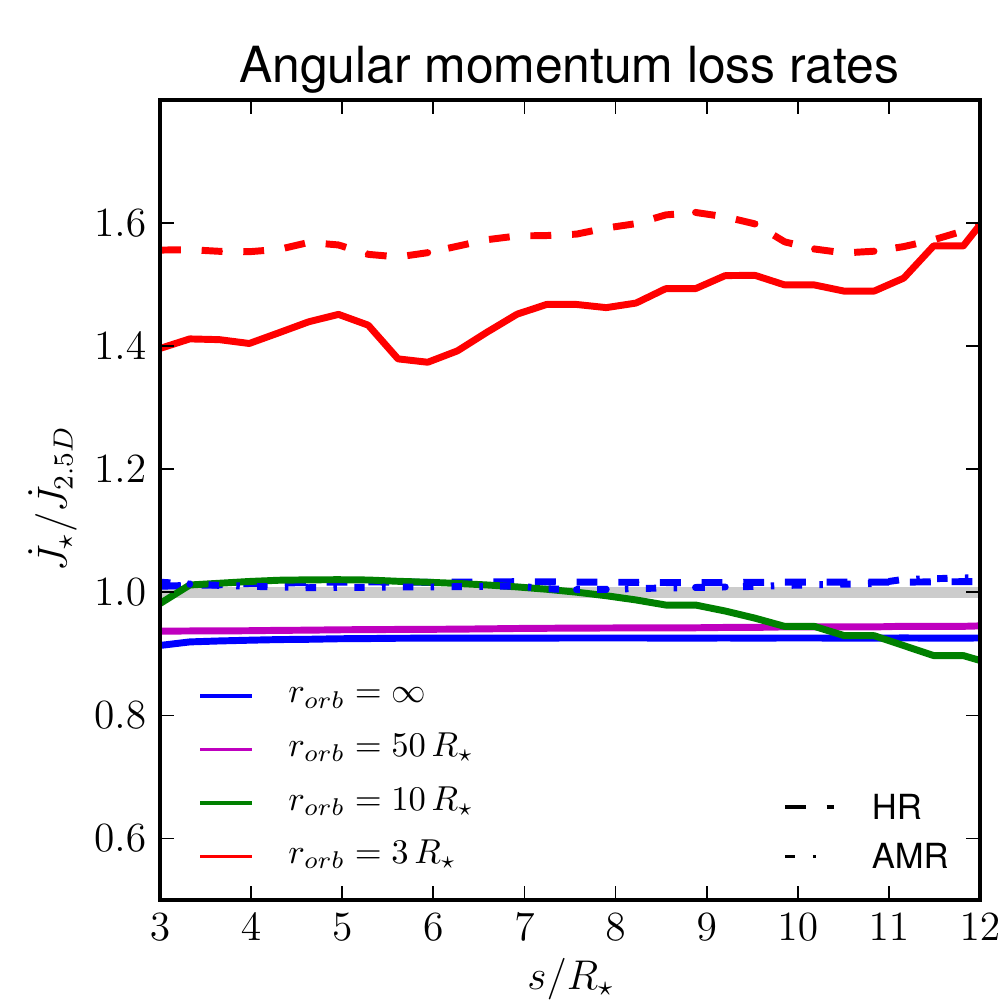}
  \caption{Mass and angular momentum loss rates as a function of the
    integration box [$s$] averaged over a few stellar
    rotations. The loss rates are normalized to the
    loss rates obtained from an equivalent 2.5D axisymmetric model
    \citep[see][]{Strugarek:2014uh}. The fiducial resolution cases are
    shown in solid line, the 'high' resolution (HR) cases in dashed
    lines and the AMR case in dash-dotted line. The various rotating
    frames are labeled with different colors.}
  \label{fig:mdot_jdot}
\end{figure}

We first assess the effect of the grid resolution and of the rotating
frame on the integrated properties of the stellar wind. We define the
mass and angular momentum loss rates due to the wind by
\begin{eqnarray}
  \label{eq:mass_loss_rate}
\dot{M}_{\star} &=& \oint{\rho \mathbf{v} \cdot  d\mathbf{A}} \, , \\
  \label{eq:angmom_flux}
\dot{J}_\star &=& \oint{ \varpi \left( v_\phi -
    B_\phi\frac{\mathbf{v}_p\cdot\mathbf{B}_p}{\rho|\mathbf{v}_p|^2}
  \right) \rho\mathbf{v}\cdot d\mathbf{A}}\, ,
\end{eqnarray}
where $\oint\,d\mathbf{A}$ represents the integral over a
two-dimensional, closed surface.
When a steady-state is reached, integrals
(\ref{eq:mass_loss_rate}-\ref{eq:angmom_flux}) can be \textit{in principle} equivalently
evaluated on any surface enclosing the star. For instance we show in Figure
\ref{fig:mdot_jdot} the loss rates computed with integrals over
cubes of size $2\,s$ centered on the star ($\dot{M}_{\star}$ is shown
in the left panel, and $\dot{J}_{\star}$ in the right panel). The loss
rates are normalized to loss rates obtained from a 2.5D
axisymmetric simulation \citep[see][]{Reville:2014ud,Strugarek:2014uh}
with a resolution equivalent (in 2D) to the HR cases. We
immediately see that the integrals are, in most of the cases, constant
functions of $s$, indicating that a steady-state is reached and that
mass and angular momentum are conserved in the flow.  

The cases with no rotating frame are shown in blue (the solid line
represents the fiducial resolution, the dashed line the 'high'
resolution--HR and the dashed-dotted line the AMR case). The mass
and angular momentum loss rates in the HR cases
differ by less than 2\% from the reference 2.5D simulation. The
fiducial resolution cases differs from $\sim$ 10\% from the HR cases,
which is a simple consequence of
the very coarse resolution that was used in those cases. When a
slowly rotating frame is added ($r_{orb}=50\, R_{\star}$, magenta
lines), only a marginal difference is observed in both loss rates. 

We observe than the mass loss rates are mostly unaffected by the
rotating frame: on the left panel each style of curve (solid and
dashed) differ from less than 2\% from one another. The angular
momentum loss rate (right panel) is nonetheless severely 
altered when a rotating frame is added. The curves are even non-constant (cases with
$r_{orb} \le 10\, R_{\star}$ in green and red) which is a due to the
difficulty to get a steady-state for cartesian grids with high
rotation rates $\Omega_{0}$. Higher resolution (dashed red line) seems
to help getting rid of those numerical issues, although in the case of
$r_{orb}=3\, R_{\star}$ the HR resolution should still be increased to
adequately model the stellar wind and obtain a constant angular
momentum loss rate consistent with the cases with no rotating frame. 

\subsection{Conservation properties}
\label{sec:rotating-frame-1}

Using the cylindrical coordinates $(\varpi,\varphi,z)$, and under the
assumption of axisymmetry, five ideal-MHD
quantities conserved along each magnetic field line can be defined by
\citep[see, \textit{e.g.},][]{Lovelace:1986kd,Ustyugova:1999ig}
\begin{eqnarray}
  \label{eq:Chiprime_2D}
  K(\psi)  &\equiv& \rho\frac{\mathbf{v}_p\cdot\mathbf{B}_p}{|\mathbf{B}_p|^2} \, , \\
  \label{eq:Lambda_2D}
  \Lambda(\psi) &\equiv& \varpi \left( v_\varphi -B_\varphi\frac{B_p}{\rho
      v_p} \right) = \varpi\left(v_\varphi - \frac{B_\varphi}{K}\right) \, , \\
  \label{eq:Omega_2D}
  \Omega_{e}(\psi) &\equiv& \frac{1}{\varpi}\left(v_\varphi
  -\frac{v_p}{B_p}B_\varphi \right) = \frac{1}{\varpi}\left(v_\varphi - \frac{KB_\varphi}{\rho}\right)\, , \\
  \label{eq:S_2D}
  S(\psi) &\equiv& P\, \rho^{-\gamma} \, , \\
  \label{eq:E_2D}
  E(\psi) &\equiv& \frac{1}{2}\left(\mathbf{v}_p^2 -v_\varphi^2\right)+
  \frac{\gamma}{\gamma-1}\rho^{\gamma-1} S - \frac{GM_\star}{r}
  + v_\varphi B_\varphi\frac{K}{\rho}\, ,
\end{eqnarray}
where $\psi$ is a magnetic field line label and the subscript '$p$'
stands for the poloidal component of a vector field. Our initial and
boundary conditions do not introduce \textit{a priori} any
non-axisymmetry (except maybe at the outer boundary). These five quantities
should hence be conserved with
a perfect model. The non-conservation can only arise from numerical
errors, in our case principally due to the use of a cartesian grid
which is not well adapted to the spherical geometry of the problem.
In order to asses quantitatively the conservation properties of our 3D
model, we compute on each three-dimensional field line the relative deviation from the
field-line averaged conserved quantity $Q$, defined by
\begin{equation}
  \label{eq:fl_dev}
  \bar{Q} \equiv \left|\frac{Q-\langle Q \rangle_{fl}}{\langle Q \rangle_{fl}}\right| \, ,
\end{equation}
where $\langle\rangle_{fl}$ stands for the average over one
three-dimensional field line. We sample the surface of the star with $20$ points in latitude
and $3$ points in longitude as seed points of magnetic field lines. We
obtain in each cases approximately the same number of closed and open
field lines. 

\begin{figure}[htbp]
  \centering
  \includegraphics[width=\linewidth]{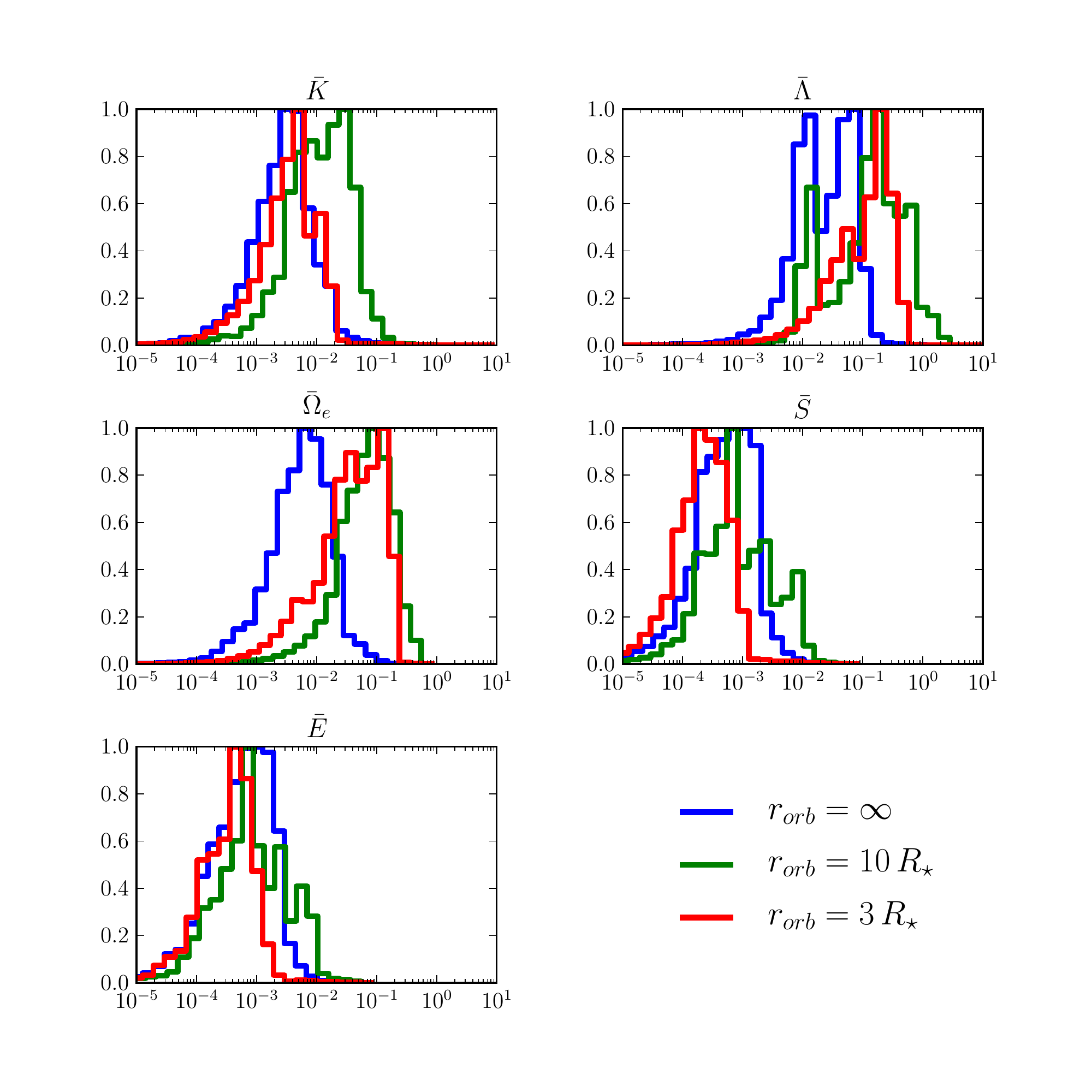}
  \caption{Normalized probability distribution functions of the relative
    deviations of the conserved quantities
    (\ref{eq:Chiprime_2D}-\ref{eq:E_2D}) for open field lines. Case 1
    is in blue ($r_{orb}=\infty$), case 5 in green ($r_{orb}= 10\,
    R_{\star}$) and case 6 in red ($r_{orb}=3\, R_{\star}$). The three
    cases were run with the fiducial resolution (see table \ref{tab:tab_cases}).}
  \label{fig:prop_conserv}
\end{figure}
 
We show in Figure \ref{fig:prop_conserv} the normalized probability
density function (PDF) of the relative deviation of conserved quantities for
cases 1, 5, and 6. We restricted our analysis to the open field
lines region, where the mass
and angular momentum of the star are extracted by the wind. The PDFs of $\bar{K}$, $\bar{S}$, and
$\bar{E}$ peak well below one and do not extend significantly above
10\%. These three quantities can be considered, as a result, to be well conserved by our
model. The PDFs of $\bar{\Lambda}$ and $\bar{\Omega}_{e}$ peak closer
to 1 (above 10\%) in the cases with a rotating frame (green and red). 
This is another way to see the numerical difficulty that is
naturally imposed by our cartesian grid. The degree of
non-conservation is sufficiently high so that the wind models with
$r_{orb}=3\, R_{\star}$ and $r_{orb}= 10\, R_{\star}$ cannot be used
reliably to, \textit{e.g.}, derive general trends about stellar
wind torques and mass loss rates. Note nevertheless that, at first
glance, the solution $r_{orb}=3\, R_{star}$ seemed perfectly well
behaved (see Figure \ref{fig:3D}). The estimation of the angular
momentum loss rate based on an integral over this regular Alfv\'en
surface could not, in this case, provide an accurate calculation because of
the lack of conservation of $\Lambda$ we just highlighted.

\begin{table}[htbp]
  \caption{Statistics of the relative deviations of the conserved
    quantities (\ref{eq:Chiprime_2D}-\ref{eq:E_2D}) for open field lines.}
  \label{tab:tab_stats}
  \centering
\begin{tabular}{l|c|rrrr}
\multicolumn{2}{}{} &             Mean &    Standard Dev. &        Skewness &        Kurtosis \\
\hline
\multirow{5}{*}{$r_{orb}=\infty$ }
                             & $\bar{K}$        & $7.0 \, 10^{-3}$ & $2.5 \, 10^{-2}$ & $2.3 \, 10^{1}$ & $6.9 \, 10^{2}$ \\
                             & $\bar{\Lambda}$  & $4.9 \, 10^{-2}$ & $5.0 \, 10^{-1}$ & $1.2 \, 10^{2}$ & $1.5 \, 10^{4}$ \\
                             & $\bar{\Omega}_e$ & $1.1 \, 10^{-2}$ & $1.5 \, 10^{-2}$ &         $ 6.0 $ & $6.6 \, 10^{1}$ \\
                             & $\bar{S}$        & $9.8 \, 10^{-4}$ & $1.3 \, 10^{-3}$ &         $ 6.1 $ & $7.0 \, 10^{1}$ \\
                             & $\bar{E}$        & $1.2 \, 10^{-3}$ & $1.5 \, 10^{-3}$ &         $ 6.7 $ & $8.6 \, 10^{1}$ \\
\hline
\multirow{5}{*}{$r_{orb}=\infty$ (HR)}
                             & $\bar{K}$        & $8.6 \, 10^{-3}$ & $1.7 \, 10^{-2}$ & $1.2 \, 10^{1}$ & $2.3 \, 10^{2}$ \\
                             & $\bar{\Lambda}$  & $3.6 \, 10^{-2}$ & $5.2 \, 10^{-2}$ & $1.0 \, 10^{1}$ & $1.9 \, 10^{2}$ \\
                             & $\bar{\Omega}_e$ & $1.0 \, 10^{-2}$ & $1.5 \, 10^{-2}$ &         $ 4.1 $ & $2.7 \, 10^{1}$ \\
                             & $\bar{S}$        & $1.3 \, 10^{-3}$ & $2.1 \, 10^{-3}$ &         $ 8.2 $ & $1.2 \, 10^{2}$ \\
                             & $\bar{E}$        & $1.5 \, 10^{-3}$ & $2.2 \, 10^{-3}$ &         $ 7.7 $ & $1.1 \, 10^{2}$ \\
\hline
\multirow{5}{*}{$r_{orb}=50\,R_{\star}$}
                             & $\bar{K}$        & $4.4 \, 10^{-3}$ & $2.5 \, 10^{-2}$ & $2.5 \, 10^{1}$ & $7.9 \, 10^{2}$ \\
                             & $\bar{\Lambda}$  & $4.6 \, 10^{-2}$ & $5.4 \, 10^{-1}$ & $1.1 \, 10^{2}$ & $1.4 \, 10^{4}$ \\
                             & $\bar{\Omega}_e$ & $3.7 \, 10^{-3}$ & $6.6 \, 10^{-3}$ &         $ 8.3 $ & $1.4 \, 10^{2}$ \\
                             & $\bar{S}$        & $3.2 \, 10^{-4}$ & $7.9 \, 10^{-4}$ & $1.3 \, 10^{1}$ & $2.4 \, 10^{2}$ \\
                             & $\bar{E}$        & $3.3 \, 10^{-4}$ & $9.4 \, 10^{-4}$ & $1.6 \, 10^{1}$ & $3.4 \, 10^{2}$ \\
\hline
\multirow{5}{*}{$r_{orb}=10\,R_{\star}$}
                             & $\bar{K}$        & $2.4 \, 10^{-2}$ & $3.4 \, 10^{-2}$ &         $ 9.8 $ & $2.0 \, 10^{2}$ \\
                             & $\bar{\Lambda}$  & $3.1 \, 10^{-1}$ & $6.9 \, 10^{-1}$ & $6.1 \, 10^{1}$ & $5.9 \, 10^{3}$ \\
                             & $\bar{\Omega}_e$ & $1.1 \, 10^{-1}$ & $1.0 \, 10^{-1}$ &         $ 1.7 $ &         $ 3.5 $ \\
                             & $\bar{S}$        & $2.4 \, 10^{-3}$ & $3.5 \, 10^{-3}$ &         $ 3.4 $ & $2.1 \, 10^{1}$ \\
                             & $\bar{E}$        & $2.3 \, 10^{-3}$ & $3.9 \, 10^{-3}$ &         $ 5.6 $ & $5.4 \, 10^{1}$ \\
\hline
\multirow{5}{*}{$r_{orb}=3\,R_{\star}$}
                             & $\bar{K}$        & $7.6 \, 10^{-3}$ & $2.5 \, 10^{-2}$ & $2.3 \, 10^{1}$ & $6.8 \, 10^{2}$ \\
                             & $\bar{\Lambda}$  & $2.2 \, 10^{-1}$ &          $ 5.7 $ & $1.4 \, 10^{2}$ & $2.1 \, 10^{4}$ \\
                             & $\bar{\Omega}_e$ & $7.2 \, 10^{-2}$ & $6.5 \, 10^{-2}$ &         $ 1.1 $ &         $ 1.2 $ \\
                             & $\bar{S}$        & $4.4 \, 10^{-4}$ & $1.2 \, 10^{-3}$ & $1.8 \, 10^{1}$ & $4.1 \, 10^{2}$ \\
                             & $\bar{E}$        & $6.4 \, 10^{-4}$ & $1.5 \, 10^{-3}$ & $1.9 \, 10^{1}$ & $4.8 \, 10^{2}$ \\
\hline
\multirow{5}{*}{$r_{orb}=3\,R_{\star}$ (HR)}
                            & $\bar{K}$        & $5.3 \, 10^{-3}$ & $8.0 \, 10^{-3}$ & $2.0 \, 10^{1}$ & $6.2 \, 10^{2}$ \\
                            & $\bar{\Lambda}$  & $4.3 \, 10^{-2}$ & $3.7 \, 10^{-2}$ &         $ 1.7 $ &         $ 5.2 $ \\
                            & $\bar{\Omega}_e$ & $2.8 \, 10^{-2}$ & $2.7 \, 10^{-2}$ &         $ 1.8 $ &         $ 4.9 $ \\
                            & $\bar{S}$        & $3.1 \, 10^{-4}$ & $5.5 \, 10^{-4}$ & $1.3 \, 10^{1}$ & $3.1 \, 10^{2}$ \\
                            & $\bar{E}$        & $3.7 \, 10^{-4}$ & $6.9 \, 10^{-4}$ & $1.9 \, 10^{1}$ & $5.2 \, 10^{2}$ \\
\hline
\end{tabular}
\end{table}

We give more extensive statistical properties of
the distributions of deviations in the open field lines region in
table \ref{tab:tab_stats} for all the cases listed in table
\ref{tab:tab_cases}. It immediately appears that in all cases, the
mean deviation (and its standard deviation) is the highest for
$\bar{\Lambda}$ and $\bar{\Omega}_{e}$. The HR cases bring a
significant improvement in the conservation of those two quantities,
and in particular in their standard deviation. This shows that with a
sufficiently high resolution, the angular momentum loss rate
calculation could be robustly estimated from such three-dimensional
models. The lack of conservation of
$\Lambda$ in stellar wind models is also generally accompanied by
large longitudinal variations of the rotation rate of the wind.
In the context of magnetic star-planet interactions, the rotation of
the wind is naturally key to assess the eventual effect on the
planetary magnetosphere and on the secular evolution of the
system. As a consequence, only stellar wind models with acceptable
conservation properties should be used to assess the effects of those interactions.




\section{Conclusions}
\label{sec:conclusions}

We have presented a comparative study of simple, 3D models of the
stellar wind of cool stars. We focused our study on the numerical
problems that can arise from the use of \textit{(i)} a cartesian grid
and \textit{(ii)} a (fast) rotating frame in the context of
star-planet interactions. 

Our results suggest that, without sufficient spatial resolution, a rotating
frame with a rotation rate higher than the stellar rotation rate
should be avoided. The numerical experiments presented
here were conducted for a small stellar rotation rate. Because of this
small rotation rate, small errors arising from the cartesian
grid can lead to dramatic changes in the stellar
wind solution. We expect the issues encountered in this work to be
significantly lower in cases with higher stellar rotation rates,
and adaptative mesh refinement seems to be an adequate, generic solution to
overcome those numerical difficulties.

\acknowledgments{
AS thanks T. Matsakos for discussions about the modelling of star-planet systems
in 3D. This work was supported by the ANR 2011 Blanc 
\href{http://ipag.osug.fr/Anr\_Toupies/}{Toupies}
and the ERC project 
\href{http://www.stars2.eu/}{STARS2} (207430). The
authors acknowledge CNRS INSU/PNST and CNES/Solar Orbiter
fundings. AS acknowledges support from the Canada's Natural Sciences and
Engineering Research Council and from the Canadian Institute of
Theoretical Astrophysics (National fellow). We acknowledge access to supercomputers
through GENCI (project 1623), Prace, and ComputeCanada infrastructures.
}

\normalsize


\end{document}